\documentclass{jfm}
\usepackage{graphicx}
\usepackage{epstopdf, epsfig}
\usepackage{amsmath}
\usepackage{xcolor,colortbl}
\usepackage{multirow}
\usepackage{longtable}
\usepackage{array}
\usepackage{hhline}

\definecolor{Grey}{gray}{0.9}

\shorttitle{Effects of moment of inertia on rising spheres}
\shortauthor{J. B. Will, D. Krug}

\title{Dynamics of freely rising spheres: the effect of moment of inertia}

\author{Jelle B. Will\corresp{\email{j.b.will@utwente.nl}}\aff{1}
    \and Dominik Krug\corresp{\email{d.j.krug@utwente.nl}}\aff{1}}

\affiliation{\aff{1}Physics of Fluids Group, Max Planck UT Center for Complex Fluid Dynamics, Faculty of Science and Technology, MESA+ Institute, and J.M. Burgers Centre for Fluid Dynamics, University of Twente, P.O. Box 217, 7500 AE Enschede, The Netherlands}

\begin{document}

\maketitle

\begin{abstract}
The goal of this study is to elucidate the effect the particle moment of inertia (MOI) has on the dynamics of spherical particles rising in a quiescent and turbulent fluid. 
To this end, we performed experiments with varying density ratios $\Gamma$, the ratio of the particle density and fluid density, ranging from $0.37$ up to $0.97$. At each $\Gamma$ the MOI was varied by shifting mass between the shell and the center of the particle to vary $I^*$ (the particle MOI normalised by the MOI of particle with the same weight and a uniform mass distribution). Helical paths are observed for low, and `3D chaotic' trajectories at higher values of $\Gamma$. The present data suggests no influence of $I^*$ on the critical value for this transition $0.42<\Gamma_{\textrm{crit}}<0.52$.
For the `3D chaotic' rise mode we identify trends of decreasing particle drag coefficient ($C_d$) and amplitude of oscillation with increasing $I^*$. Due to limited data it remains unclear if a similar dependence exists in the helical regime as well. 
Path oscillations remain finite for all cases studied and no `rectilinear' mode is encountered, which may be the consequence of allowing for a longer transient distance in the present compared to earlier work. Rotational dynamics did not vary significantly between quiescent and turbulent surroundings, indicating that these are predominantly wake driven.
\end{abstract}

\begin{keywords}
\end{keywords}

\section{Introduction}
It is widely known that freely rising spheres can exhibit a host of different and complex path oscillations. Numerous studies have been devoted to this topic, which is of interest e.g. as a paradigmatic case for fluid-structure interactions.
Canonically, the independent parameters considered are the density ratio $\Gamma = \rho_p/\rho_f$ and the particle Reynolds number $Re = \langle v_z \rangle D/\nu$ (or a related quantity such as the Galileo number $Ga = \sqrt{|1-\Gamma|gD^3}/\nu$). Here, $\rho_p$ and $\rho_f$ denote the particle and fluid densities, respectively, $\langle \cdot \rangle$ indicates a time or ensemble average and $v_i$ is the velocity component of the particle velocity $\boldsymbol{v}$ in direction\,$i$ (which in the definition of the Galileo number is replaced by the buoyancy velocity $V_b = \sqrt{|1-\Gamma|gD}$). Further, $D$ is the sphere diameter, $\nu$ the kinematic viscosity of the fluid, and $g$ is the acceleration due to gravity. Both parameters, $\Gamma$ and $Re$, are related to the vertical momentum balance
\begin{equation}
	\Gamma \dfrac{\textrm{d} \boldsymbol{v}}{\textrm{d} t} = \dfrac{\boldsymbol{F}_f(Re)}{m_f} + (1-\Gamma)g\boldsymbol{e}_z, \label{eq:MoItranslation}
\end{equation}
where $\boldsymbol{F}_f$ is the fluid forcing on the body, $m_f$ is the particle mass, and $\boldsymbol{e}_z$ is a unit vector pointing opposite to the direction of gravity.

The Reynolds number dependence enters implicitly in (\ref{eq:MoItranslation}) via the fluid forcing ${\boldsymbol{F}_f}$ on the sphere. Once $Re \gtrapprox 200$ \citep{Jenny2003}, vortex shedding sets in in the particle wake, which results in an approximately periodic forcing and a complex dynamical coupling between particle motion and the surrounding flow field \citep{bearman1984,parkinson1989,williamson2004vortex,govardhan_2005}. The most comprehensive investigation of the $\Gamma$-$Re$ parameter space reported to date is by \citet{Horowitz:2010}. Based on their experiments, these authors conclude that a critical density ratio $\Gamma_{crit}$ exists, which governs the onset of path oscillations. The value of $\Gamma_{crit}$ was shown to exhibit a $Re$ dependence and path oscillations did not occur for $\Gamma_{crit} > 0.36$ for $260 < Re < 1550$ and $\Gamma_{crit} > 0.6$ at $Re > 1550$. \citet{Horowitz:2010} also state that the presence of path oscillations is associated with a high drag regime, for which the values of the drag coefficient $C_d$ significantly exceed values reported for a fixed sphere at similar $Re$. However, there remain fundamental and largely unexplained discrepancies in the literature on the topic. This is most evident in the spread of reported $C_d$  values (see figure\,\ref{fig:MoICd}\,({\it b\/}) and the corresponding discussion), but also manifests in differences in the reported rise modes. Whereas \citet{Horowitz:2010} reported only planar ('zigzaging') trajectories, other studies find helical or spiralling motions \citep{preukschat1962,shafrir1965,Auguste2018,will2020kinematics,will2020rising} for comparable parameter values. Also the `rectilinear mode' described in \citet{Horowitz:2010} for $\Gamma >\Gamma_{crit}$, in which particles rise straight without path oscillations, is not observed consistently elsewhere; e.g. \citet{preukschat1962} report a reduction in oscillating amplitude with increasing $\Gamma$ within this regime, but did not encounter perfectly 'vertical' (i.e. non-oscillating) trajectories.  

There certainly are a host of possible explanations for these differences and the origin of some of them may well be linked to the precise experimental conditions. The latter include the precision of the particle fabrication, residual disturbances in the flow, and the size of the tank among potentially many more. However, recent findings \citep{namkoong2008,Mathai:2017,Mathai2018,will2020rising} also suggest a more systematic cause as they point to an additional relevance of the rotational dynamics in setting the overall particle dynamics. Rotations of the sphere are governed by
\begin{equation}
	I^*\Gamma \dfrac{\textrm{d} \boldsymbol{\omega}}{\textrm{d} t} = \dfrac{10 \boldsymbol{T}_f}{m_f D^2},\label{eq:eom_rot}
\end{equation}
where, $\boldsymbol{\omega}$ is the angular velocity of the sphere and $\boldsymbol{T}_f$ the torque induced by the fluid. The additional independent parameter introduced by (\ref{eq:eom_rot}) is the dimensionless moment of inertia (MOI), $I^* = I_p/I_\Gamma$ with $I_\Gamma = \pi/60 \rho_p D^5$ the moment of inertia of a particle with a uniform material density of $\rho_p$. Note that the definition of $I^*$ is chosen such that its value is entirely determined by the mass distribution within the particle and independent of $\Gamma$ (i.e. fluid properties). For a homogeneous sphere $I^* = 1$ and $I^* <1$ ($I^* > 1$) if the mass is accumulated towards (away from) the centre. Note, however, that the dynamically relevant parameter implied by (\ref{eq:eom_rot}) is given by the product $I^* \Gamma$. 

While there is no explicit coupling between (\ref{eq:MoItranslation}) and (\ref{eq:eom_rot}), the two degrees of freedom can interact via the flow field, e.g. through a Magnus force $\boldsymbol{F}_m \sim \boldsymbol{\omega} \times \boldsymbol{v}$. 
A potential relevance of the MOI as an additional parameter was already mentioned by \citet{ryskin_leal_1984}. For the case of cylinders, its importance has been established via systematic studies in two-dimensional simulations \citep{namkoong2008,Mathai:2017}. 
More recently, \citet{Mathai2018} also uncovered a regime transition induced by a variation in the MoI  for spheres rising in a turbulent flow. These authors also reported differences when the particles were rising in still fluid, but these observations remained qualitative and limited to two different values of $I^*$ at a single density ratio ($\Gamma$). 
Moreover, recent experiments by \citet{will2020rising} confirmed the general relevance of rotational dynamics for rising or settling spheres. By introducing a center of mass offset, these authors selectively varied the rotational dynamics while keeping $\Gamma$ and $Ga$ constant. This led to a resonant behaviour between particle rotation and wake shedding with significant impact on parameters such as oscillation amplitude and $C_d$. Further and remarkably, also horizontal path oscillations ceased almost fully once the offset got large enough to mostly suppress rotational motion.

On this basis, it is the goal of this study to systematically explore the effect variations in $I^*$ have on the rise behaviour of light spheres. Little can be gleaned from existing data sets (most of them based on particles with a non-uniform mass distribution) to answer this question, as this parameter is generally not reported. Therefore, we designed and manufactured particles to perform new experiments exploring the parameter space systematically. Details on this can be found in \S\,\ref{sec:ExpSetup}. Afterwards, we present and discuss the results of the experiments in \S\,\ref{sec:Results}. Additionally, we investigate the effects of background disturbances in the fluid and the effect of the time between experiments (waiting time) (see  \S\,\ref{sec:waiting_time}) and finally conclude in \S\,\ref{sec:Conclusion}.

\section{Experimental setup and procedures}\label{sec:ExpSetup}
\subsection{Particle design and manufacture}
\begin{figure}
	\centerline{\includegraphics[width=1.0\textwidth]{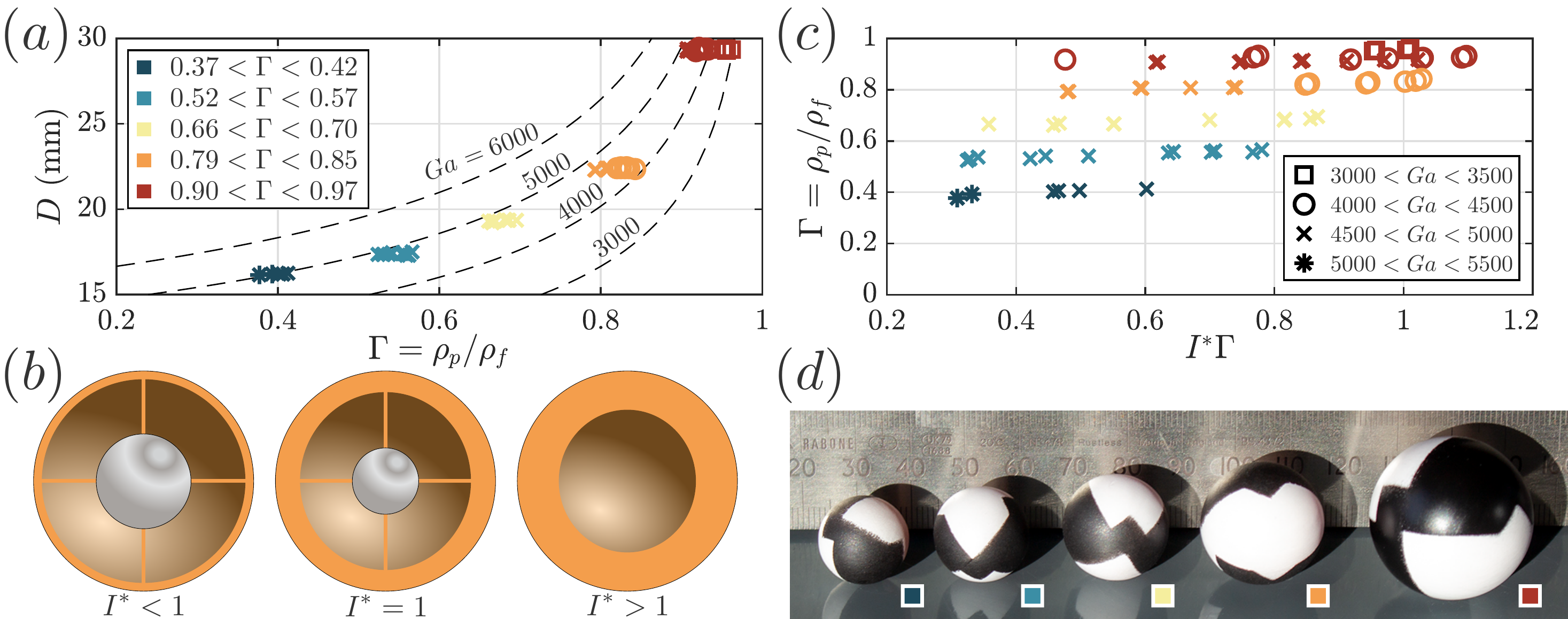}}
	\caption{({\it a\/}) Schematic of the particle design showing how we vary the moment of inertia of a particle by placing metal bearing ball of varying sizes in the centre while keeping the density nominally constant. ({\it b\/}) Picture showing the finished particles, one for each density ratio range (indicated by the coloured squares). ({\it c\/}) Particle diameter ($D$) and density ratio ($\Gamma$) of all particles used in the experiments. Isolines of $Ga$ show the variation in this parameter. ({\it d\/}) Particle density ratio $\Gamma$ and the dimensionless moment of inertia $I^* \Gamma$ for all particles. }
	\label{fig:particle_properties}
\end{figure}
We aim to vary the MOI while keeping $Ga$ and $\Gamma$ nominally constant. This is achieved by shifting the weight between the outer shell of the particle and a metal ball of varying size at its center as required, see figure\,\ref{fig:particle_properties}\,({\it a\/}).  The shells are designed using 3D CAD software and 3D printed on a RapidShape 30L printer with a horizontal resolution of 21 $\mu$m and a layer thickness of 25 $\mu$m. The print is performed in two halves, which are then glued together and sanded to smoothen the surface. In a last step, a pattern is painted on the particle to enable the rotation tracking, resulting in the final particles shown in figure\,\ref{fig:particle_properties}\,({\it b\/}). The mass of the paint and glue contributed less than $0.5\%$ of the total particle mass and hence does not induce a significant centre of mass offset nor variation of $I^*$.
Measured values of the final particle weight and diameter are used to update the CAD model in order to obtain an more accurate value of the MOI.
The final particles were smooth to the touch and we expect that the residual surface roughness will not affect the outcome of the experiments significantly. This is based on the fact that in the present range of $Re$, the flow over the roughness elements is laminar \citep{Achenbach:1972} and the skin friction will thus not depend on the roughness height \citep{Moody1944}.

The sphericity achieved with this method is high with diameter measurements (taken with a caliper) differing by less than $1\%$ of the diameter at different cross sections. An overview of the parameter space covered in this study is shown in  figure\,\ref{fig:particle_properties}\,({\it c,d\/}). Throughout this work the marker colour will designate the different $\Gamma$ regimes and the different marker types indicate the value of $Ga$. Line color is used to indicate ranges of $I^*\Gamma$. For each $\Gamma$-regime, $I^*$ is varied as much as physically possible, the resulting ranges are shown in figure\,\ref{fig:particle_properties}\,({\it d\/}) in terms of $I^*\Gamma$.

\subsection{Experimental setup and methods}
\begin{figure}
	\centerline{\includegraphics[width=1.0\textwidth]{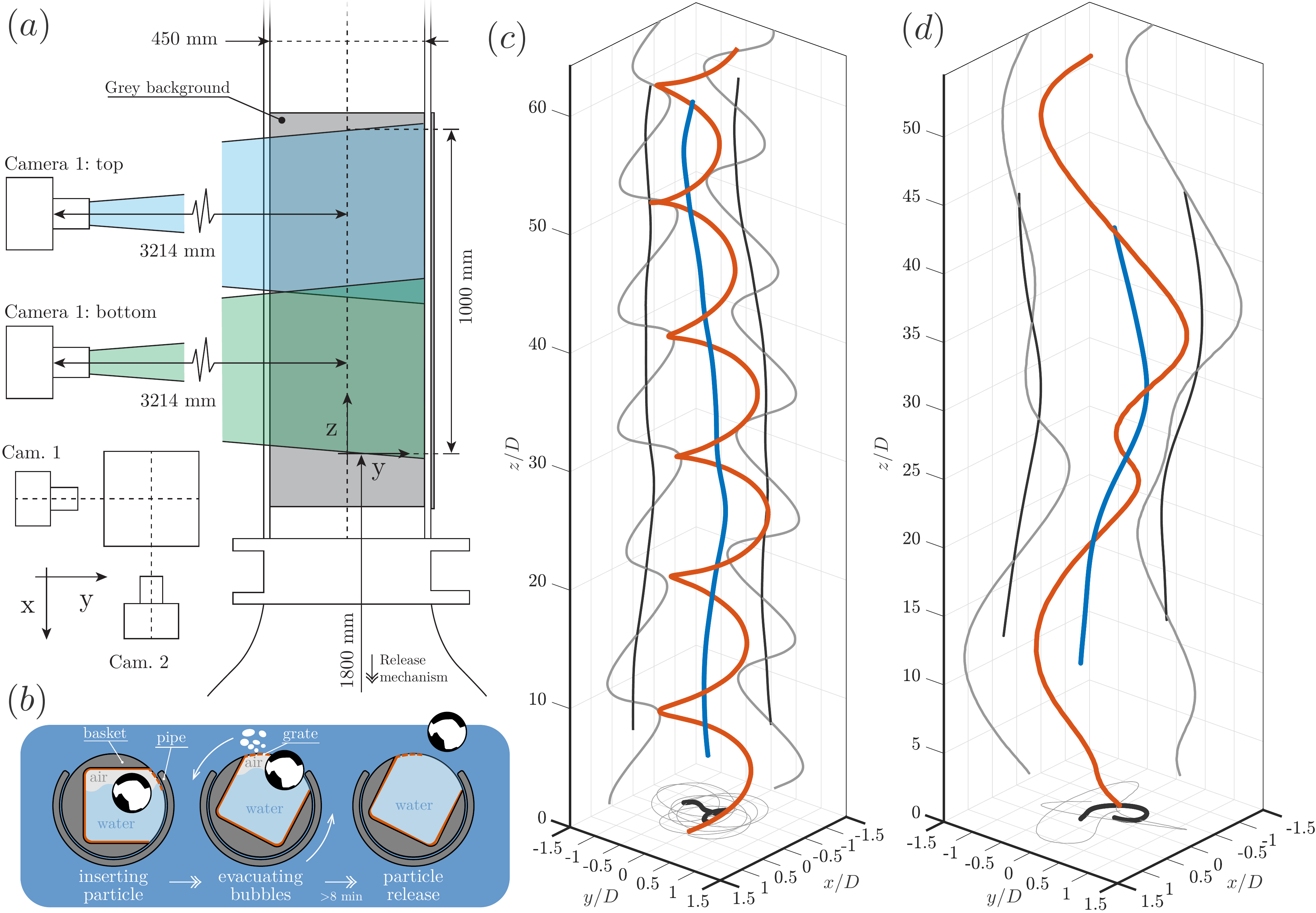}}
	\caption{({\it a\/}) Schematic of the test section of the Twente Water Tunnel along with the camera setup. ({\it b\/}) Schematic detailing the release mechanism and procedure. Left: particle is inserted in the basket and pushed to the centre of the tank. Centre: basket is rotated in the pipe to expose the grate and to allow the air to escape the basket. Right: after waiting for at least 8 minutes, the particle is released by rotating the basket slightly further. ({\it c\/}) An example of a particle trajectory (red) and its center-line (blue). Also shown are projections of the particle path onto the sides and bottom of the domain. Note that horizontal axes are re-scaled with respect to the vertical to highlight the path oscillations. The properties of this particle are: $\Gamma = 0402$, $I^* = 1.140$, $D = 16.2$ mm. ({\it d\/})  An example of a particle trajectory for a secondary particle with properties: $\Gamma = 0.666$, $I^* = 0.827$, $D = 19.3$ mm.}
	\label{fig:ExpSetup}
\end{figure}
All experiments were performed in the approximately 3 m high, water filled, test section of the Twente Water Tunnel facility. The setup is schematically shown in figure \ref{fig:ExpSetup}\,({\it a\/}). The lab is temperature controlled at $20^\circ$C, thus we assume constant fluid properties $\rho_f = 998$ kg m$^{-3}$ and $\nu = 1.0035 \times 10^{-6}$ m$^2$ s. The particles were released using a specifically built release mechanism located approximately 1.8 m below the measurement region. The release mechanism, depicted in figure \ref{fig:ExpSetup}\,({\it b\/}), consists of a pipe with a cutout from which the particle can be released. Particles can be inserted into this pipe without draining the tank using a basket. Once pushed to a cutout in the pipe in the center of the tank, this basket can be turned, thereby releasing the trapped air through a grate while keeping the particle inside. After the bubbles have risen to the top of the tank, the water is left to settle for at least 8 minutes before the particle is released by gently tilting the basket further. Doing so did not cause significant rotation of the spheres upon leaving the release mechanism. In \S\ref{sec:waiting_time}, we validate the dependence of the behaviour on the waiting time, since this was previously found to be critical to the rise behaviour \citep{Horowitz:2010}.

Once the particle has entered the measurement section, the particle is recorded by two pairs of perpendicularly placed high-speed cameras (PHOTRON Fastcam AX200 with 1024×1024 pixels at 256 grey levels, fitted with ZEISS Milvus 100mm lenses). The cameras are placed more than 3 meters away from the centre of the tunnel in order to get a near isometric view of the pattern on the particles. Stacking two camera pairs (see figure \ref{fig:ExpSetup}\,({\it a\/})) allows tracking of the particles over a distance exceeding 1 meter in the vertical direction. Grey background panels are used to contrast with the white and black pattern on the surface of the particles. The frame rate of the cameras was adjusted depending on the rise-velocity in order to keep the inter-frame translation between 2-6 pixels.

Based on the calibration of the camera position and taking into account parallax effects, the 3D particle position in space was reconstructed from the two orthogonal views. The origin of the coordinate system is located at the base of the measurement domain in the centre of the tunnel. The directions are defined as depicted in figure \ref{fig:ExpSetup}\,({\it a\/}), with $x$ and $y$ spanning the horizontal plane and $z$ pointing upward, i.e.\ opposite to the direction of gravity. The obtained position data is smoothed by convolution with a Gaussian kernel to obtain the trajectories shown as red curves in figures \ref{fig:ExpSetup}\,({\it c,\,d\/}). We further obtain the velocity and acceleration of the particle by convolution with the derivatives of a Gaussian kernel \citep{Mordant:2004}. The window sizes and standard deviations of the kernels were varied based on the particle size and were determined to minimize the noise, while leaving the underlying signal intact \citep{Mathai:2016}.

Additionally, we track the orientation of the spheres by matching the particle images to rendered projections of the patterns at different orientations \citep{will2020kinematics}. With the sequence of orientations known, the angular velocity and acceleration can be derived as described in \citet{will2020rising}.

From the raw data for the trajectories we then determined the dominant frequency of oscillation by considering the horizontal velocity signals. This result is then used to determine the centre-line of the trajectory indicated by the blue curves in figures \ref{fig:ExpSetup}\,({\it c,\,d\/}) \citep{will2020kinematics}. The amplitude of the path oscillations ($\hat{a}$) is based on the distance between the trajectory and its centre-line, i.e.\ the trajectory data are drift-corrected. 

Finally, in \S\ref{sec:f2t} we perform experiments with a mean downward flow present and active grid generated turbulence in the same facility. In these measurement we only use the top two cameras and the measurement region is 0.8m downstream of the active grid. In these experiments we release the particle using the same mechanism but with a downward flow in the channel, balancing out the particle rise velocity. The mean flow velocity in the channel is measured using a magnetic flow meter and is kept constant during the experiment. The active turbulence grid at the top of the channel is turned on and when the particle is in the measurement domain the recording is started. The particle can stay in this region for a long time producing recordings in access of a duration of 30 seconds. Typical measurement times easily exceeded 30s, resulting in very good statistics for this configuration. The setup is identical to that used by \citet{Mathai2018} with $Re_\lambda \approx 300$ and $\Xi = D/\eta \approx 100$, with $\eta$ the Kolmogorov length scale of the turbulent flow.

\section{Results and discussion}\label{sec:Results}
\begin{figure}
	\centerline{\includegraphics[width=1.0\textwidth]{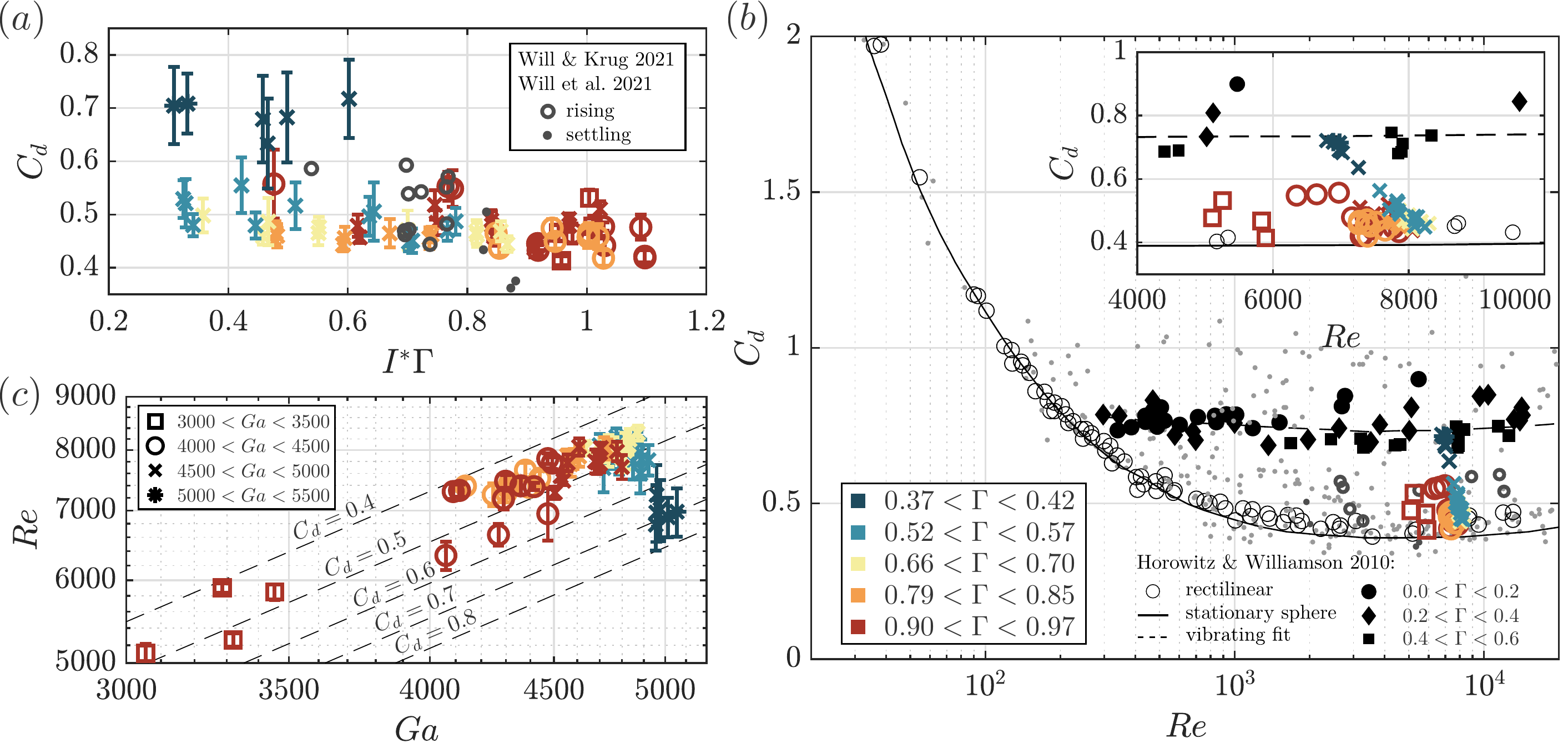}}
	\caption{({\it a\/}) Drag coefficient as a function of the dimensionless moment of inertia $I^*\Gamma$. In addition to present results, we plot data from \citet{will2020kinematics,will2020rising} for isotropic rising and settling spheres. ({\it b\/}) A comparison of the drag coefficient with compiled data from literature versus the particle Reynolds number. 	The inset shows a more detailed view of the current data and the results for a stationary sphere (solid black line) and the ``vibrating fit'' from \citet{Horowitz:2010}. ({\it c\/}) Particle Reynolds number vs. Galileo number; diagonal dashed lines indicate constant drag coefficient. In all panels, the symbols correspond to ranges in particle Galileo number and the colours indicate ranges in density ratio $\Gamma$}
	\label{fig:MoICd}
\end{figure}

\subsection{Particle drag coefficient}\label{sec:Drag}

We start by considering the drag coefficient
\begin{equation}
	C_d = \dfrac{4(1-\Gamma)D g}{3\langle v_z \rangle^2} = \dfrac{4}{3}\dfrac{Ga^2}{\langle Re \rangle^2},\label{eq:CdGaRe}
\end{equation}
as a function of the dimensionless moment of inertia $I^*\Gamma$ in figure \ref{fig:MoICd}\,({\it a\/}). Most saliently, these results cluster into a high-drag regime ($C_d \approx 0.7$) and a low-drag regime with $C_d \approx 0.45$. Albeit not as pronounced as the difference between these regimes, there is further a distinct trend of decreasing $C_d$ with increasing $I^* \Gamma$ within the low-drag regime. In figure\,\ref{fig:MoICd}\,({\it a\/}) we also included relevant data from \citet{will2020kinematics,will2020rising} for both rising and settling particles. These data points are largely in line with the low-drag mode in the present data set. 
Only for the lowest values of $\Gamma$ considered here ($0.37 < \Gamma < 0.42$) do we encounter the high-drag regime. Given the limited data points, no conclusions on the dependence of $C_d$ on $I^*$ can be drawn in this case. Finally, we also note that the transition between the two drag regimes appears independent of $I^*$ in the present data.

To establish how the observed trends ---and in particular the dependence of $C_d$ on $I^* \Gamma$---  relate to literature data, we compare our data to published values of $C_d$ \citep{preukschat1962,maccready1964,shafrir1965,kuwabara1983,Jenny2004,karamanev1996,stringham1969,veldhuis2007,veldhuis2009,allen1900,liebster1927,lunnon1928,boillat1981,will2020kinematics,will2020rising} in figure\,\ref{fig:MoICd}\,({\it b\/}). The existence of two different drag states as a function of $\Gamma$ has been observed and documented before by \citet{Horowitz:2010}. They attributed the regimes to a transition from  a ``vibrating mode'', with large path oscillations at low $\Gamma$, to a rectilinear mode with almost no path oscillations.
The respective $C_d$ values for these two regimes largely match our results (see also inset in figure\,\ref{fig:MoICd}\,({\it a\/})). However,  the threshold density ratio in the present data ($0.42 \leq \Gamma \leq  0.52$) is significantly lower compared to the critical value of $\Gamma = 0.61$ determined in \citet{Horowitz:2010} (for unknown $I^*$). Regarding the dependence of $C_d$ on  $I^*$, it is noteworthy that the $C_d$ values of the rectilinear mode ---and at the same time also those for a stationary sphere--- are best matched at high $I^* \Gamma$. The increase in $C_d$ with decreasing  $I^* \Gamma$ then leads to a deviation from these reference data. Some part, but certainly not all of the spread in the drag data reported in the literature, might therefore indeed be attributed to differences in the moment of inertia. However, the effect appears less strong compared to the $\Gamma$-dependence, which remains the dominant parameter in governing the particle drag. 

Finally, we also document the particle Reynolds number defined as $Re = \langle v_z \rangle_n D/\nu$. These results are shown in figure \ref{fig:MoICd}\,({\it c\/}) as a function of $Ga$. The $Re$-range for all our experiments are in the Newtonian drag regime, $Re \gtrapprox 1000$ \citep{clift1971}, for which the drag is pressure dominated ($C_d \sim D^2$) and mostly independent of $Re$. The Reynolds and Galileo numbers are linked to the particle drag coefficient as shown in (\ref{eq:CdGaRe}) and indicated by diagonal dashed lines in the figure with bi-logarithmic scales. The variation in $Re$ is small between most particles and there appears to be no systematic relation between $C_d$ and $Re$. This indicates that the variation in $C_d$ observed in figure \ref{fig:MoICd}\,({\it a\/}) in the low drag mode is indeed an $I^*$ effect and not related to $Re$. In the following, we will investigate in more detail how varying $\Gamma$ and the MOI changes properties of the particle trajectories.

\subsection{Particle trajectories}\label{sec:trajectories}
\begin{figure}
	\centerline{\includegraphics[width=1.0\textwidth]{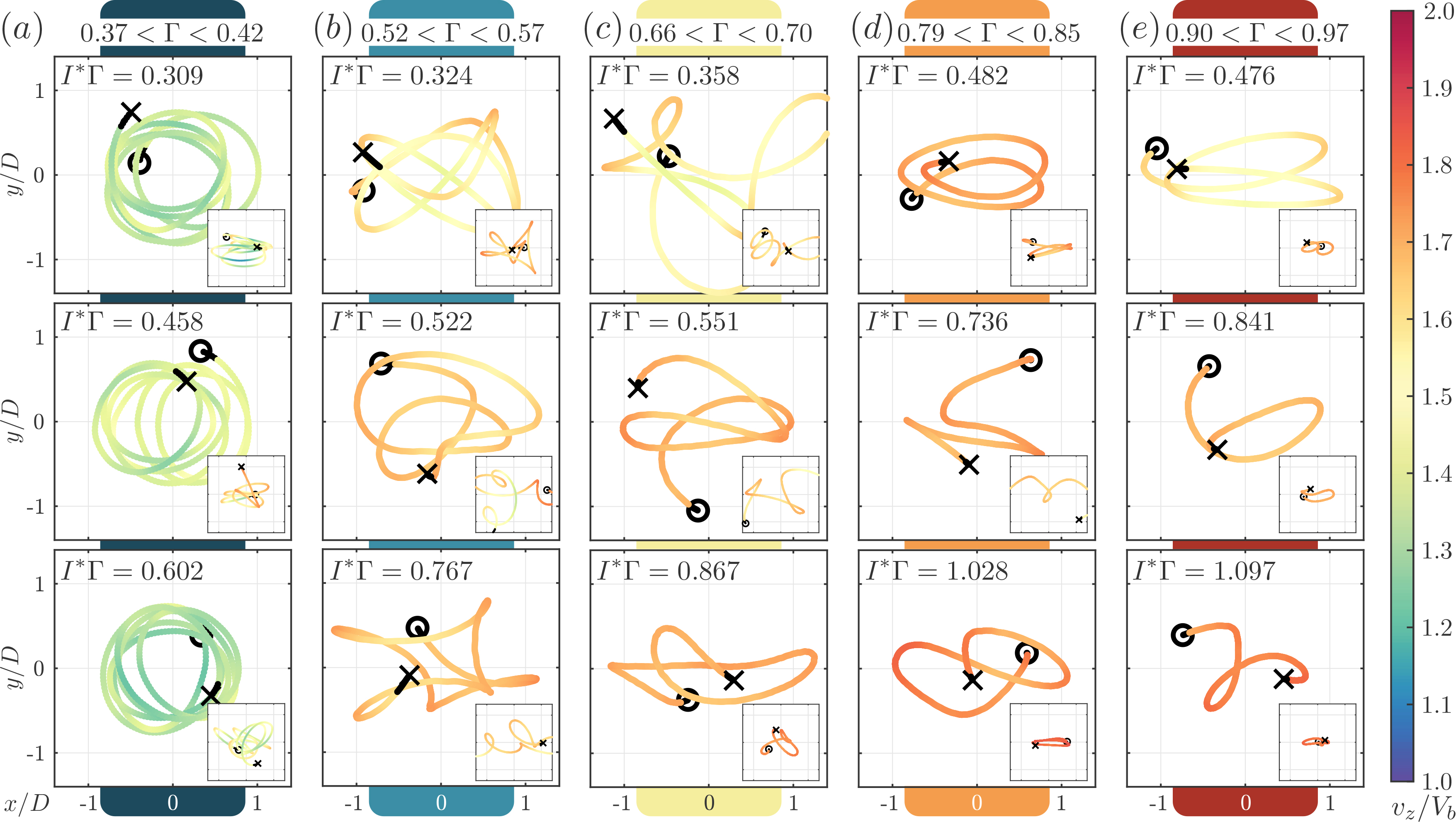}}
	\caption{Representative particle trajectories in the horizontal ($x-y$) plane. The trajectories are colour coded using the normalized vertical velocity $v_z/V_b$, symbols mark the start (circle) and end (cross) of the recorded path. $\Gamma$ increases from left to right (columns ({\it a--e\/}) and vertically in each column the value of $I^*$ increases from top to bottom. For each case two trajectories are shown: the main window shows the most frequently occurring trajectory and the inset shows the most aberrant run for that same particle.}
	\label{fig:MoItrajectories}
\end{figure}
In figure\,\ref{fig:MoItrajectories} a number of representative trajectories for the five ranges of $\Gamma$ and for three values of $I^*$ are shown. To provide a sense of the variability, the main panel shows the most commonly observed trajectory, while the inset shows a second trajectory for the same particle that is (visually) most different from the typical one. Surprisingly, the horizontal trajectories corresponding to the high drag-regime ($0.37 < \Gamma < 0.42$) are almost circular, indicating that the path is helical for these cases. This result is unlike the planar `zigzag' observed by \citet{horowitz2008,Horowitz:2010} for spheres with a low density ratio at similar $Ga$. However, helical trajectories are not uncommon and have been observed for spheres at low density ratios \citep{preukschat1962,Karamanev:1992,karamanev1996,veldhuis2009} as well as for bubbles \citep{ellingsen2001,mougin2001,mougin2006}. In the work by \citet{veldhuis2009}, these helical trajectories were connected to a different drag scaling, which could be attributed to lift-induced drag resulting from the shedding of additional vorticity in the wake. This non standard drag behaviour for spiralling trajectories was also noted in the numerical work by \citet{Auguste2018}. They found a similar spiralling regime at low $\Gamma$, but their data set is limited to $Ga \leq 700$. \citet{Karamanev:1992} state that all particles with ${Re} > 130$ and $\Gamma \leq 0.3$ rose in a spiralling trajectory. These spirals featured a constant angle pitch angle between $\boldsymbol{v}$ and $\boldsymbol{e}_z$ of about $\pm 29^\circ$. For the spiralling regime here, we find this angle to be around $18.5^\circ$ independent of $I^*$. In this context it is also important to note that trajectories became more circular for spheres with centre of mass offset in resonance with their natural frequency \citep{will2020rising}. However, this can be ruled out as a factor here since the required offset of $0.03D$ certainly exceeds our fabrication tolerance. Furthermore, it is extremely unlikely to randomly hit resonance for all particles across a range of $I^*$ values. We are therefore convinced that the present trajectories reflect the genuine particle behaviour at the present values of $Ga$, $\Gamma$ and $I^*$.

The dynamics in the low drag-regime for $\Gamma > 0.52$ are distinctly different from the spiralling motion at low density ratios.
While there remains periodicity in the sequence of direction changes, the turning angles appear random.
The resulting behaviour is characteristic of the `3D chaotic' regime \citep{Auguste2018}, which applies to all particles with $\Gamma > 0.52$ here. These results are at odds with the findings of \citet{Horowitz:2010}, who found a vertical rise regime for all particles larger than $\Gamma \approx 0.61$ at comparable $Ga$.

Finally, there is no clear trend visible in the shape of the horizontal trajectories for varying MOIs. We will proceed to investigate the horizontal motion more quantitatively in order to elucidate such effects.

\subsection{Fluctuations of the horizontal velocity}\label{sec:Vfluct}
\begin{figure}
	\centerline{\includegraphics[width=1.0\textwidth]{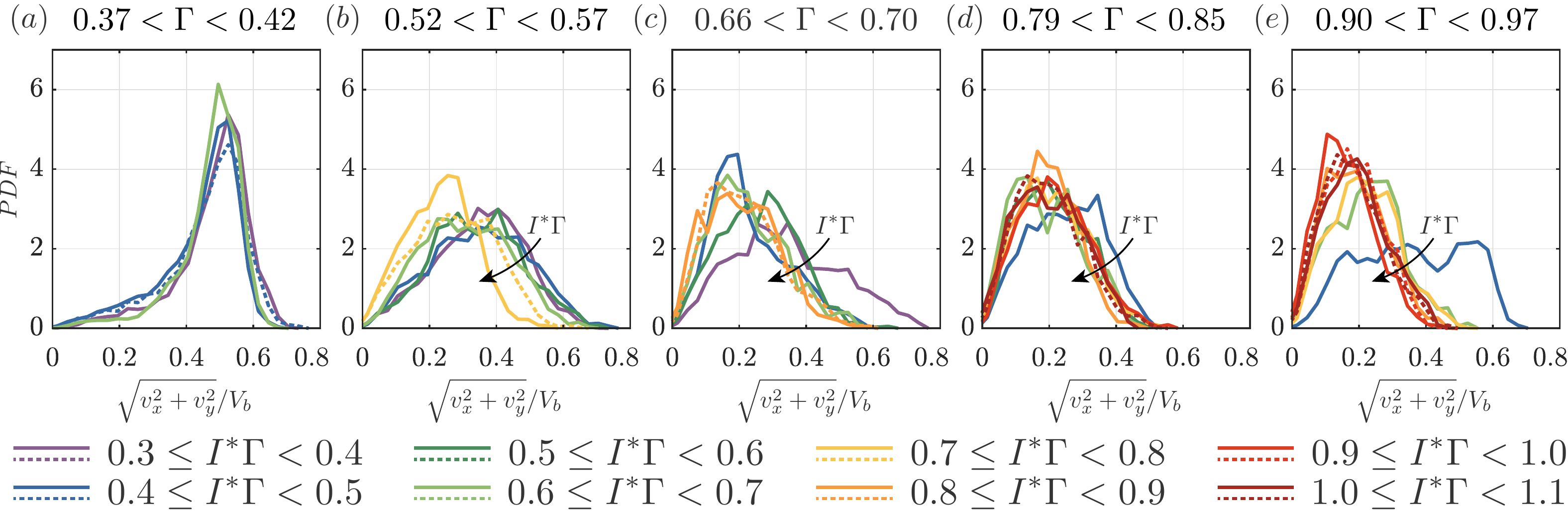}}
	\caption{ ({\it a--e\/}) Normalized probability density functions of dimensionless horizontal velocity fluctuations at different $\Gamma$. The colour coding of the lines represents ranges in $I^*\Gamma$; dashed lines are used for different combinations of $I^*$ and $\Gamma$ resulting in the same value of their product. Note: each line represents an average over multiple particles with the same nominal properties.} 
	\label{fig:vel_pdfs}
\end{figure}

To uncover the time-varying dynamics of the rising spheres, it is useful to consider the probability density functions (PDFs) of the horizontal velocity. The distributions of the normalised horizontal velocity ($\sqrt{v_x^2+v_y^2}/V_b$) are shown for various values of $I^*\Gamma$ at different $\Gamma$ in figure\,\ref{fig:vel_pdfs}\,({\it a--e\/}). As a most obvious trend, we note a change in skewness from negative for $\Gamma \leq 0.42$ (figure\,\ref{fig:vel_pdfs}\,({\it a\/})) to positive skew at  $\Gamma \geq 0.52$ (figure\,\ref{fig:vel_pdfs}\,({\it b--e\/})). This is indicative of the transition from spiralling motion, for which the horizontal velocity is generally high, to the `3D chaotic' state, for which strong horizontal translation occurs more intermittently.

The effect of varying $I^*\Gamma$ (indicated by the line colour) is negligible at $0.37 < \Gamma < 0.42$, as evidenced by figure\,\ref{fig:vel_pdfs}\,({\it a\/}). 
Given the limited (by physical constraints) range of $I^*$ at this density ratio, it remains unclear to what extent this indicates a lesser importance of the rotational dynamics in the spiralling regime (see also \S\,\ref{sec:RotDyn} on this). 

For $\Gamma \geq 0.52$ (i.e.\ in the `3D chaotic' state), however, a clear dependence of the PDFs of $\sqrt{v_x^2+v_y^2}/V_b$ on $I^*\Gamma$ emerges.  We observe that for low $I^*\Gamma$, the distribution is in general broader and extends further towards high velocities, whereas at high $I^*\Gamma$ the distributions is narrow and the peak shift towards lower velocities. This trend is most evident in figure\,\ref{fig:vel_pdfs}\,({\it c,\,e\/}), showing that at very low $I^*$ the distributions become rather flat, reminiscent of a fluttering behaviour with more pronounced horizontal motion.

It should be noted that the trends discussed here for $\sqrt{v_x^2+v_y^2}/V_b$ also manifest in the statistics of the vertical velocity $v_z$ (not shown), albeit with inverse effects regarding the skewness and the MOI dependence. While there is a distinct negative correlation between instantaneous vertical velocity and instantaneous horizontal velocity, which is also obvious from the colour coding in figure \ref{fig:MoItrajectories}\,({\it b--e\/})), the magnitude of the velocity, $||\boldsymbol{v}||$ is not constant but fluctuates quasi-periodically for all cases. 

\subsection{Oscillation frequency and amplitude}\label{sec:freq_amp}
\begin{figure}
	\centerline{\includegraphics[width=1.0\textwidth]{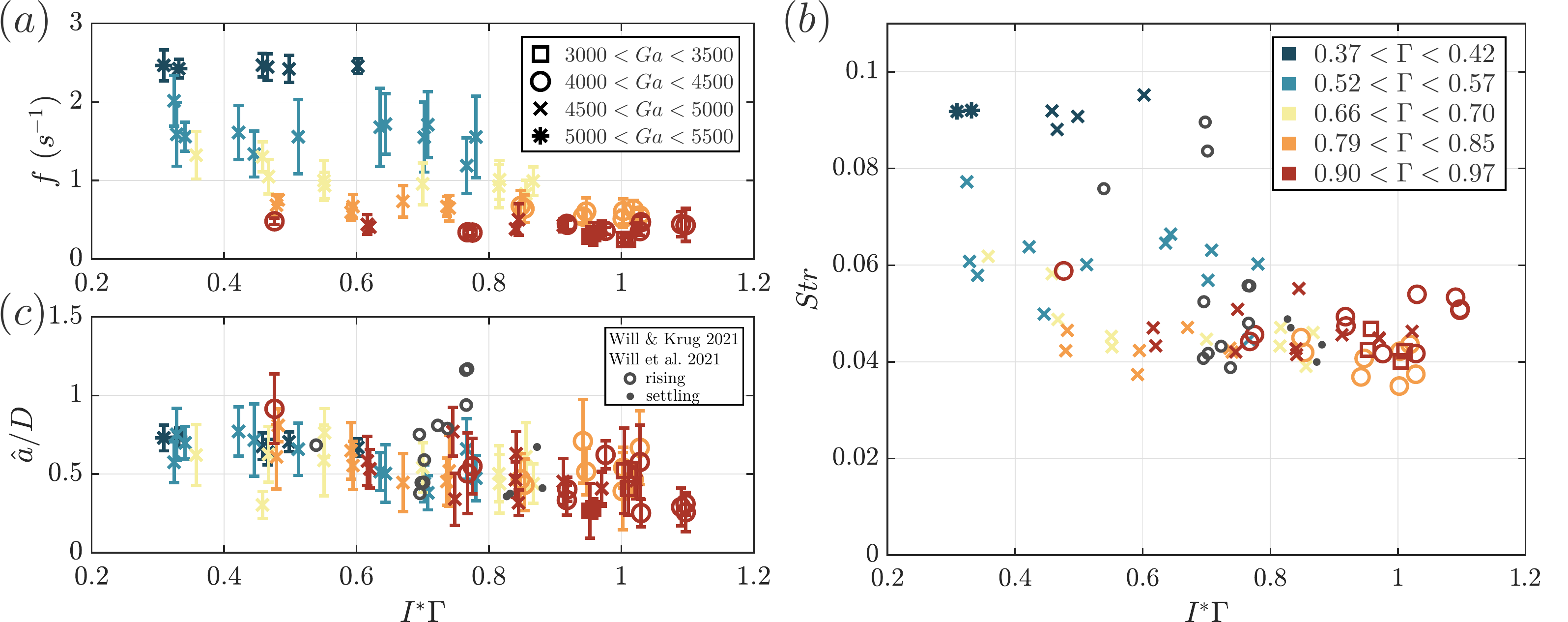}}
	\caption{({\it a\/}) Measured oscillation frequency in Hz, ({\it b\/}) Strouhal number and ({\it c\/}) amplitude of the path oscillations normalized by the particle diameter as a function of the dimensionless moment of inertia $I^*\Gamma$.}
	\label{fig:MoIfreq_amp}
\end{figure}
In all experiments performed here, we observed significant oscillations in the trajectory of the rising spheres. In this section, we will characterize these in terms of the frequency and amplitude of the horizontal oscillations. In figure\,\ref{fig:MoIfreq_amp}\,({\it a\/}), we start by plotting the frequency $f$  as a function of the dimensionless moment of inertia $I^*\Gamma$. It is evident from this figure that there is no significant dependence of $f$ on $I^*\Gamma$ as the data points with the same colour (i.e. constant $\Gamma$) are at a near constant frequency. Note that the error bars on the frequency, indicating the spread in the data, are very small for all density ratios besides $0.52 < \Gamma < 0.57$. This indicates that even though some of the motion appears quite random, there exists a strong dominant frequency associated with the vortex shedding. The outlier at $0.52 < \Gamma < 0.57$ is most likely related to this case falling within the transitional regime between helical paths and 3D chaotic patterns.

In figure \ref{fig:MoIfreq_amp}\,({\it b\/}), we show the frequency in dimensionless form in terms of the Strouhal number, defined as:
\begin{equation}
	Str = \dfrac{fD}{\langle v_z \rangle}.
\end{equation}
This normalization separates the data into two regimes, akin to those encountered for $C_d$ and for the trajectories. For $\Gamma >0.66$ we find that the Strouhal number takes a constant value of approximately $Str$ = 0.04-0.05, while for $0.37 < \Gamma < 0.42$ we find $Str \approx$ 0.09. 
Only the data for $0.52 < \Gamma < 0.57$ does not completely fall in line with this decomposition and lies at a slightly higher value of $Str \approx 0.06$, consistent with the transitional behavior of this case mentioned above. Furthermore, also the Strouhal number appears to be rather insensitive to changes in $I^*$. The value of $Str$ $\approx$ 0.09 at low $\Gamma$ matches the results by \citet{Horowitz:2010} (figure 31) closely. They find for identical $Ga$ and $\Gamma$ also $Str \approx$ 0.09, however, the sphere is zigzagging in their case instead of spiralling as observed here. The value of $Str \approx 0.04-0.05$ for $\Gamma >0.66$ is in line with the results by \citet{preukschat1962} (see fig. 19 and page 20), who for $0.582 \le \Gamma \le 0.875$ also find $Str$ as low as 0.05.

Finally, we report  the normalized amplitude of the path oscillations $\hat{a}/D$ as a function of the $I^*\Gamma$ (see figure\,\ref{fig:MoIfreq_amp}\,({\it c\/})). Despite the scatter in these data, there appears to be a consistent trend of $\hat{a}/D$ decreasing from approximately 0.8 to 0.3 with increasing $I^*\Gamma$ over the full range accessible here. Surprisingly, the values of the amplitude are congruent for both the helical regime as well as for the more chaotic trajectories. Our results imply that decreasing $I^*\Gamma$ by either changing the internal structure of the particle or the density ratio, results in larger amplitude path oscillations. In fact, it appears that a similar trend with consistent amplitudes exists in the data of \citet{Horowitz:2010} for varying $\Gamma$ in their `zigzag' regime. However, in their experiments $I^*$ is not monitored nor controlled, which renders a quantitative comparison of both datasets impossible.

\subsection{Rotational dynamics and translational coupling}\label{sec:RotDyn}
\begin{figure}
	\centerline{\includegraphics[width=1.0\textwidth]{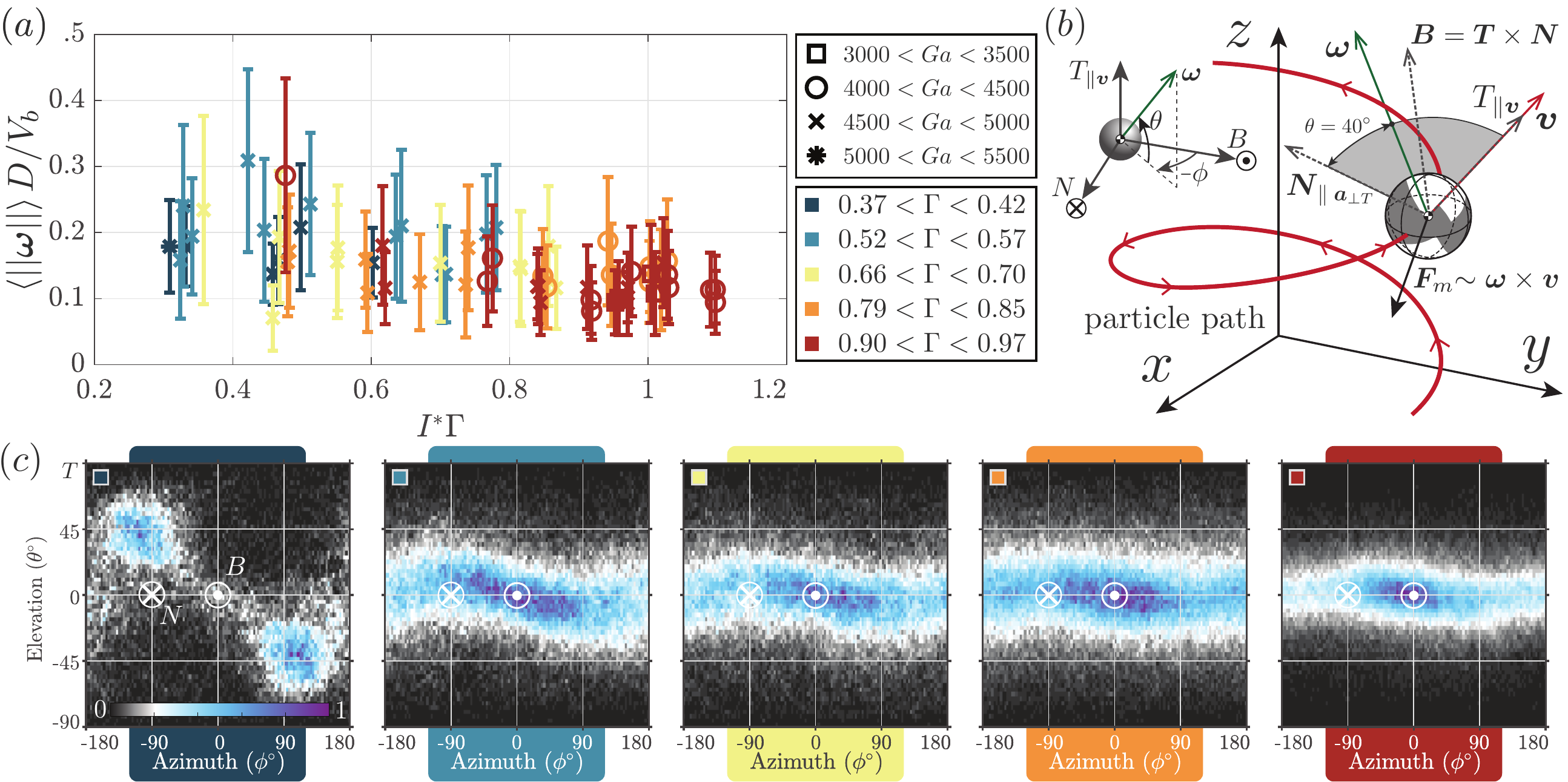}}
	\caption{({\it a\/}) Dimensionless rotation rate $\omega^* = \langle || \boldsymbol{\omega} || \rangle D / V_b$ versus the moment of inertia $I^*\Gamma$. ({\it b\/}) Frenet-Serret (TNB) coordinate system for a counter-clockwise spiralling sphere; $\boldsymbol{T}$ is parallel to the instantaneous direction of motion, $\boldsymbol{N}$ is parallel to the curvature of the path and $\boldsymbol{B}$ is defined parallel to $\boldsymbol{T} \times \boldsymbol{N}$. Definition of the direction of rotation $\boldsymbol{\omega}$ using azimuth ($\phi$) and elevation ($\theta$) with respect to the TNB coordinate system.  ({\it c\/}) Normalized histograms of the alignment of $\boldsymbol{\omega}$ in terms of $\phi$ and $\theta$ in the TNB coordinate system as a function of particle density ratio ($\Gamma$). These figures contain the data for all $I^*$-values within this $\Gamma$ range. In these figures $\theta = 90$ is aligned with $\boldsymbol{T}$, $\phi =-90$, $\theta = 0$ is aligned with $\boldsymbol{N}$, and $\phi = 0$, $\theta = 0$ is aligned with $\boldsymbol{B}$.}
	\label{fig:rr}
\end{figure}
The mechanism by which the particle moment of inertia $I^*$ affects the particle kinematics and dynamics is solely through the rotational equation of motion \ref{eq:eom_rot}, effectively scaling the particle rotation to the fluid torques. Rotational dynamics in turn affect the flow field around the body, thereby inducing a coupling with the lateral motion, e.g.\ via Magnus lift type forcing. Particle rotation can also affect vortex detachment and consequently the flow structure in the wake of the particle, an effect that is believed to be at the heart of the regime transition observed by \citet{Mathai2018}.

Therefore, the most direct parameter in investigating the impact of the MOI is the rotation rate of the body. This quantity is explored in figure \ref{fig:rr}\,({\it a\/}), where we plot the mean dimensionless rotation rate $\omega^* = \langle || \boldsymbol{\omega} || \rangle D / V_b$ versus $I^*\Gamma$. We find that the dimensionless rotation rate ($\omega^*$), similar to the drag and amplitude of the oscillation of the trajectory, shows a slight dependence on $I^*\Gamma$. The particles with lower rotational inertia indeed rotate more vigorously compared to their higher $I^*\Gamma$ counterparts. We further note that the normalization of $\langle || \boldsymbol{\omega} || \rangle$ with $D$ and $V_b$ collapses the results across all density ratios convincingly. It is important to note that the rotation rate is not affected significantly by the change from spiralling to the 3D chaotic regime.

In order to explore the importance of rotational dynamics further, we examine the alignment of the rotation vector $\boldsymbol{\omega}$ with respect to the particle acceleration along the curvature of the path. Doing so allows us to establish the relevance of Magnus lift forcing on the particle dynamics. To this end, we consider $\boldsymbol{\omega}$ in the Frenet-Serret (TNB) coordinate system \citep{zimmermann2011rotational}. As is shown in figure\,\ref{fig:rr}\,({\it b\/}), $\boldsymbol{T}$ points in  the direction of the instantaneous velocity $\boldsymbol{v}$ of the sphere, $\boldsymbol{N}$ is aligned with the curvature of the path (the acceleration of the sphere that is non-parallel to the direction of motion; $\boldsymbol{a}_{\perp \boldsymbol{T}}$), and $\boldsymbol{B} = \boldsymbol{T}\times\boldsymbol{N}$. Thus, when $\boldsymbol{\omega} \parallel \boldsymbol{B}$ the induced Magnus force $\boldsymbol{F}_m \parallel \boldsymbol{a}_{\perp \boldsymbol{T}}$, making this coordinate system very useful to study the effect of rotation on the horizontal motion. The alignment of $\boldsymbol{\omega}$ within the TNB coordinate system is given in terms of two angles: the azimuth $\phi$ and the elevation $\theta$ as indicated in figure\,\ref{fig:rr}\,({\it b\/}). The directions of $\boldsymbol{N}$ and $\boldsymbol{B}$ are also indicated in figure\,\ref{fig:rr}\,({\it d\/}), where we show normalized histograms of the alignment of $\boldsymbol{\omega}$ in the $\phi-\theta$ plane for different values of $\Gamma$. 
Most striking about these results is the enormous difference in rotational alignments between the low-$\Gamma$ and high-$\Gamma$ regimes. For the low-$\Gamma$ regime ($0.37 <\Gamma < 0.42$), we find that, depending on the direction of the spiralling motion (clockwise or counter-clockwise), the alignment of $\boldsymbol{\omega}$ is either $\theta \approx -40^\circ$ or $40^\circ$, respectively.
As illustrated in figure\,\ref{fig:rr}\,({\it b\/}) for the counter-clockwise spiralling case (but also true for clockwise rotation), this alignment implies that the Magnus lift force ($\boldsymbol{F}_m \sim \boldsymbol{\omega} \times  \boldsymbol{v}$) predominantly acts downward. This mechanism therefore leads to lift induced drag, in a manner similar to that suggested by \citet{mougin2006} and \citet{veldhuis2009}. Furthermore, with $|\phi| = 90^\circ$ the Magnus force acts perpendicular to $\boldsymbol{a}_{\perp v}$, and is hence not responsible for the lateral acceleration for the lowest $\Gamma$. Thus, the spiralling motion encountered for these particles must have a different origin, e.g.\ a rotation of the vortex shedding position relative to the direction of motion \citep{karamanev1996}. 

On the other hand, for $\Gamma >0.52$ the distribution of orientations of $\boldsymbol{\omega}$ is centred around $\boldsymbol{B}$. The histograms of all cases with $\Gamma > 0.52$ are remarkably similar underlining that these belong to the same dynamical regime. The data clusters around $\theta \approx 0$, i.e.\ $\boldsymbol{\omega}$ pointing normal to $\boldsymbol{v}$. This is consistent with vortex shedding being a main driver of particle rotation, as this induces a torque perpendicular to $\boldsymbol{T}$. Further, $\theta \approx 0$ on average implies that there is no net contribution of the Magnus force in the vertical direction. 

\begin{figure}
	\centerline{\includegraphics[width=1.0\textwidth]{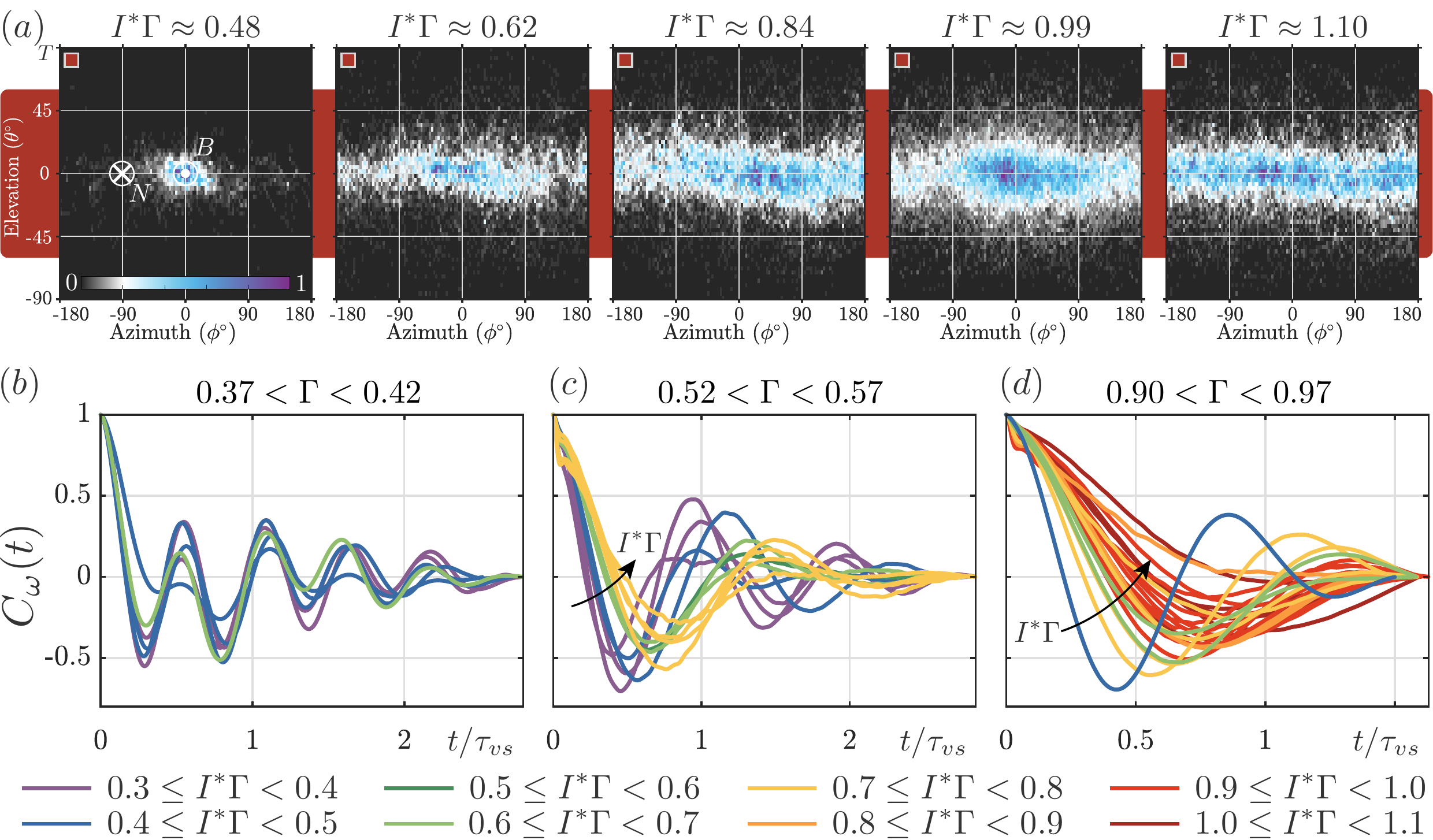}}
	\caption{ ({\it a\/}) ({\it a\/}) The effect of particle MOI on the alignment of $\boldsymbol{\omega}$ in the TNB coordinate frame. The results shown here are $0.90 < \Gamma < 0.97$ and are representative of the other density ratios in the high $\Gamma$ regime. ({\it b--d\/}) Autocorrelation functions of an (arbitrary) horizontal component of the rotation rate $\boldsymbol{\omega}$ for different density ratios. The time axis is normalized using a vortex-shedding timescale $\tau_{vs} \sim D/{\langle v_z \rangle}_n$. Each line represents results for a single particle averaged across multiple experiments; line colour indicates the value of $I^*\Gamma$.}
	\label{fig:tnb_moi}
\end{figure}

So far, we have only considered how the alignment statistics of $\boldsymbol{\omega}$ depend on $\Gamma$. Potential variations with $I^*\Gamma$ are masked by plotting data with different MOI together in figure\,\ref{fig:rr}\,({\it d\/}). To elaborate on the influence of the MOI, we present additional orientation statistics of $\boldsymbol{\omega}$  for $0.90 < \Gamma < 0.97$ at five different $I^*\Gamma$ values separately in figure\,\ref{fig:tnb_moi}\,({\it a\/}). 
At the lowest MOI ($I^*\Gamma \approx 0.48$), the alignment of $\boldsymbol{\omega}$ and $\boldsymbol{B}$ is very strong.
This behaviour is associated with a strong coupling of particle rotation and lateral acceleration resulting in a ``fluttering'' type of behaviour \citep{Mathai2018}. The alignment with $\boldsymbol{B}$ progressively weakens for higher values of $I^*$ and the distribution is approximately flat in terms of $\phi$ at $I^*\Gamma \approx 1.10$.
We note that the trend observed in figure\,\ref{fig:tnb_moi}\,({\it a\/}) is representative also for the lower $\Gamma$ cases in the 3D chaotic regime. However, the alignment with $\boldsymbol{B}$ at the respectively lowest $I*$ values is not equally as pronounced as for $0.90 < \Gamma < 0.97$. It hence appears that the 'flutter' type motion occurs at different values of $I^* \Gamma$ at different $\Gamma$.

A related analysis, that allows to more readily compare different combinations of $I^*$ and $\Gamma$, is to evaluate the autocorrelation $C_\omega$ of an arbitrary horizontal component of $\boldsymbol{\omega}$ as was done previously in \citet{Mathai2018}.
Such results are shown for 3 different ranges of the density ratio in figure\,\ref{fig:tnb_moi}\,({\it b--d\/}), where the correlation coefficient is plotted against the dimensionless time lags $t/\tau_{vs}$, with the vortex-shedding timescale $\tau_{vs} \sim D/\langle v_z \rangle_n$.
The influence of the MOI seen in figure\,\ref{fig:tnb_moi}\,({\it a\/}) is reflected also in the corresponding results in figure\,\ref{fig:tnb_moi}\,({\it d\/}). In this case, it manifests itself by a gradual transition from a periodic behaviour of $C_\omega$ (low $I^*\Gamma$, blue line) to a slow decorrelation without significant oscillation at large $I^*\Gamma$ (red lines). Note, however, that the drop to 0 at the largest $t/\tau_{vs}$ shown is an artefact of the finite observation time in all cases. A similar influence of the MOI as seen at $\Gamma \approx 0.9$ is also found at the lowest $\Gamma$ in the 3D chaotic regime in figure\,\ref{fig:tnb_moi}\,({\it c\/}). Here, the correlation timescale outgrows the vortex shedding time scale  $\tau_{vs}$ at larger $I^*\Gamma$, while at the same time the correlation also drops in magnitude.
For the lowest values of $\Gamma$ (figure\,\ref{fig:tnb_moi}\,({\it b\/})), for which spiralling trajectories are observed, the periodicity in  $C_\omega$ is most pronounced of all the cases shown. Typical periods are about $0.5\ t/\tau_{vs}$ with no obvious dependence on $I^*\Gamma$. 
Non-periodic rotations are observed most prominently at high $\Gamma$ and high MOI, which is consistent with the mostly random orientation of the $\boldsymbol{\omega}$ vector in the TNB coordinate system (see figure\,\ref{fig:tnb_moi}\,({\it a\/}) for these cases. Such behaviour was previously only reported in turbulent flow \citep{Mathai2018} and we will study this aspect in more detail in the next section. 

\section{On the effect of turbulence}\label{sec:f2t}

Previously, \citet{Mathai2018} investigated the effect of MOI on spheres rising in a downward turbulent flow. They showed that depending on their MOI the trajectories of spheres with $Ga \approx 6000$ and $\Gamma \approx 0.88$ differed significantly: For $I^*\Gamma  = 0.6$ a `zig-zag' (or `flutter') motion was observed without significant horizontal drift, whereas for $I^*\Gamma  = 1$, the behaviour changed to a tumbling motion featuring a strong mean drift.

To test if such a transition can also be observed for the present set of particles, we conducted measurements in turbulent flow. The turbulence properties $Re_\lambda \approx 300$ and $\Xi = D/\eta \approx 100$, with $\eta$ the Kolmogorov length scale, were chosen to match those of \citet{Mathai2018}. Further, we restricted measurements to two particles, both with $\Gamma \approx 0.9$ and $I^*\Gamma = 0.52$ and $I^*\Gamma =1.18$, respectively, again matching the literature conditions closely.

\begin{figure}
	\centerline{\includegraphics[width=1.0\textwidth]{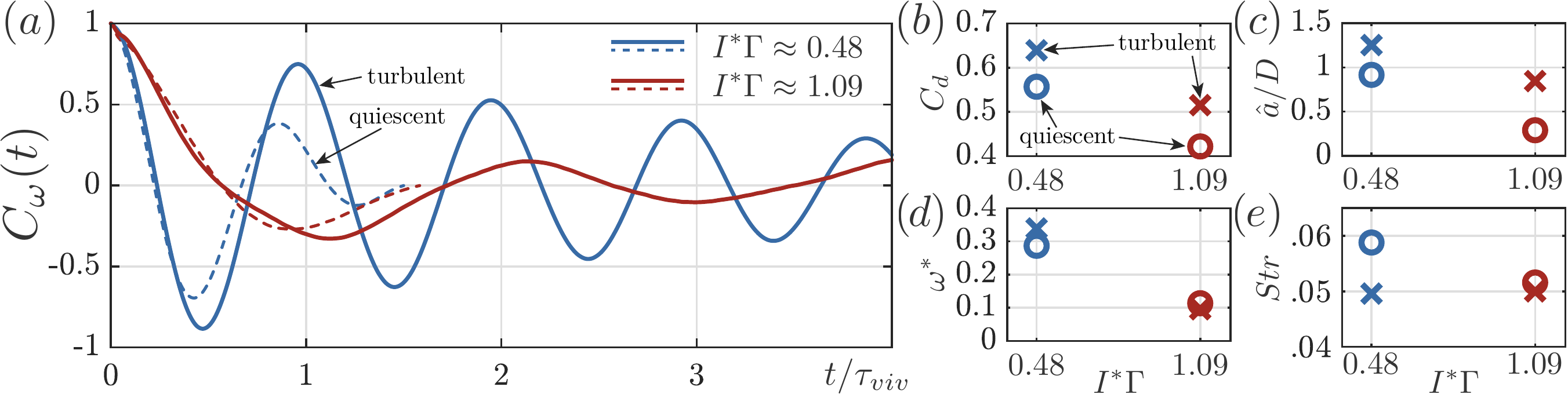}}
	\caption{({\it a\/}) Results for $C_\omega$ for $I^*\Gamma$ = 0.52 (blue) and 1.18 (red) in turbulent flow (solid lines) and quiescent surroundings (dashed lines). Comparison between the quiescent flow (circles) and turbulent flow (crosses) cases are shown in terms of drag coefficient ({\it b\/}), amplitude of the path oscillations ({\it c\/}), dimensionless rotation rate  ({\it d\/}) and Strouhal number ({\it e\/}).}
	\label{fig:autocorrelation_turbulence}
\end{figure}

Clearly and consistently with \citet{Mathai2018}, the two particles behave differently in the turbulent flow as evidenced by the results for $C_\omega$ in figure\,\ref{fig:autocorrelation_turbulence}\,({\it a\/}). However, discounting the differences caused by limited observation times in the still fluid, there appears to be little difference for the same particle in either quiescent or turbulent surroundings. This holds for $C_\omega$, but also for the fact that the tumbling behaviour reported for the turbulence measurements in \citet{Mathai2018} was not observed here. Consequently, we also see no difference in the mean drift rate of the two particles (29.07 mm/s and 26.67 mm/s for low and high MOI, respectively). It therefore appears that whether or not a sphere tumbles in turbulence might depend quite sensitively on the flow or particle parameters (e.g. values of $D/\eta$ differ by about 10\% between the two studies). Other MOI related effects, such as the decorrelation of the rotation rate and arguably the overall rise pattern, appear largely independent of the flow state.  

Naturally, small differences between the turbulent and the quiescent case arise for specific parameters and some of these are documented in figures\,\ref{fig:autocorrelation_turbulence}\,({\it b--e\/}). The drag coefficient $C_d$ (figure \,\ref{fig:autocorrelation_turbulence}\,({\it b\/})) is approximately 20\% higher in turbulence for both particles. Similarly, also the typical oscillation amplitude is higher in turbulent surroundings (figure\,\ref{fig:autocorrelation_turbulence}\,({\it c\/})). Interestingly,  the mean rotation rate (see figure\,\ref{fig:autocorrelation_turbulence}\,({\it d\/})) is barely affected by the ambient flow, which may indicate that these dynamics remain wake driven even in turbulence. For all these quantities the trend that higher values are observed at lower $I^* \Gamma$ is also preserved in turbulence. The only exception to this might be the Strouhal number in figure\,\ref{fig:autocorrelation_turbulence}\,({\it e\/}), for which no MOI dependence is visible for the turbulence results. However, with a standard deviation of 0.1 the spread in $Str$ is much larger for these data compared to those from the quiescent experiments. The difference at a given $I^*\Gamma$ is therefore only of the order of the typical fluctuations of $Str$ in turbulence. 

\section{The effect of disturbances in fluid}\label{sec:waiting_time}
\begin{figure}
	\centerline{\includegraphics[width=1.0\textwidth]{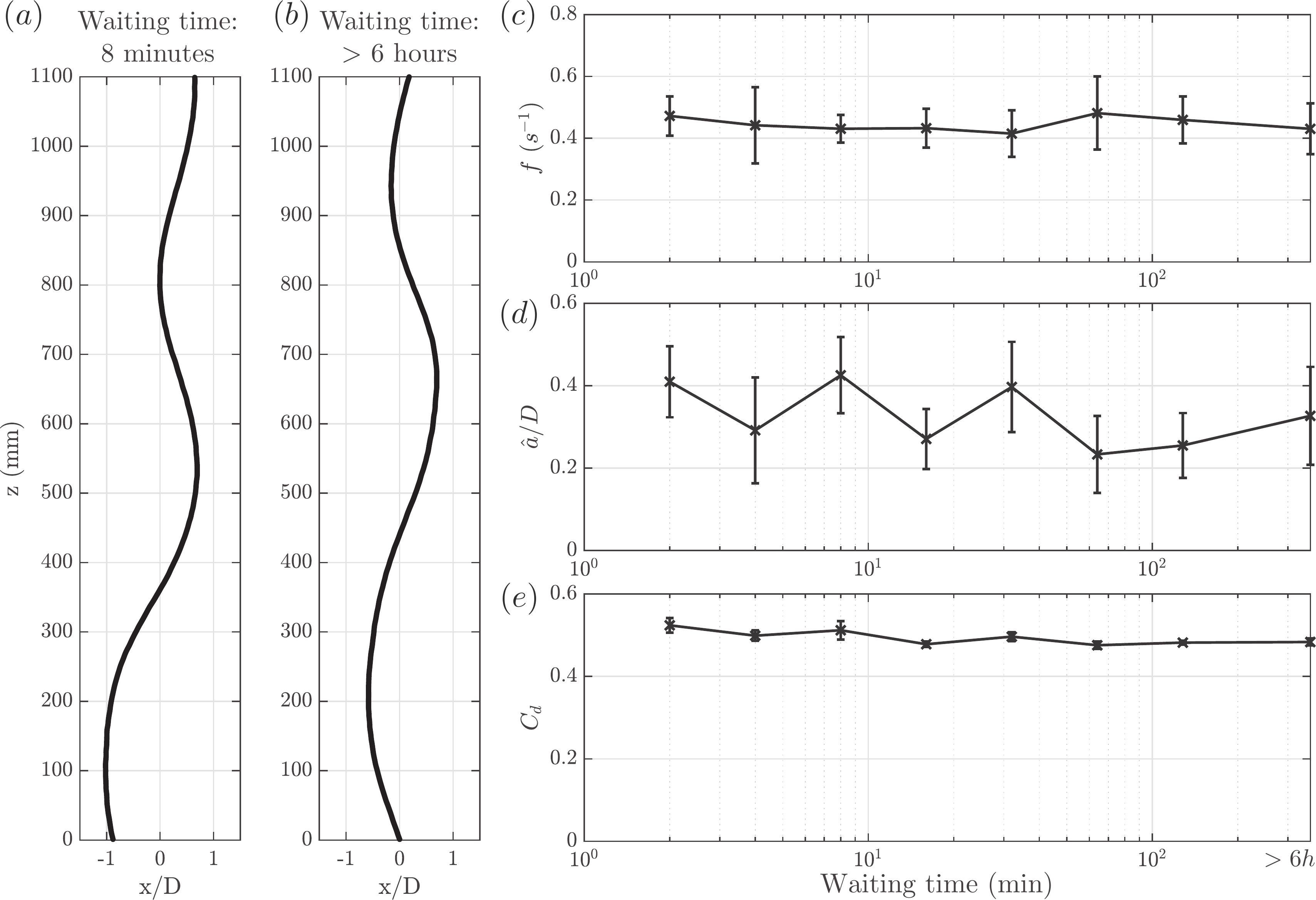}}
	\caption{Experiments performed to investigate the effect of residual fluid motion in the tank on the dynamics and kinematics of rising spheres. All results in this figure are obtained using one and the same sphere with the following properties: $\Gamma = 0.907$, $Ga = 4799$ and $I^*=0.822$. Trajectories of the same sphere as seen from the side after waiting 8 minutes ({\it a\/}) and more than 6 hours ({\it b\/}). ({\it c\/}) Average oscillation frequency as a function of the settling time. ({\it d\/}) Average oscillation amplitude as a function of the settling time. ({\it e\/}) Average drag coefficient as a function of the settling time.}
	\label{fig:settling_time}
\end{figure}
Apart from the effect of artificially introduced (turbulent) velocity fluctuations, another  point of interest is the effect of residual background disturbances in the nominally quiescent fluid. This point was stressed in the work by \citet{Horowitz:2010}. They found that the emergence of the rectilinear regime was very sensitive to the flow conditions. Any remaining disturbances in the tank caused by the insertion of the release mechanism, or caused by previous experiments, was sufficient to result in path oscillations. Since this rectilinear regime was not observed here in the parameter range specified by \citet{Horowitz:2010}, it is therefore imperative to check if this might be related to flow disturbances.

For this purpose, we performed a series of experiments varying the waiting time between the evacuation of the bubbles from the release mechanism (figure \ref{fig:ExpSetup}\,({\it b\/})) and the particle release. All measurements were performed with a single particle with properties ($\Gamma = 0.907$, $Ga = 4800$ and $I^*=0.822$) well within the rectilinear regime of \citet{Horowitz:2010}.
The waiting times considered were 2, 4, 8, 16, 32, 64, 128 and $>360$ minutes with 5 repeats performed for each of these cases.  None of these experiments displayed a rectilinear rise behaviour, and the influence of the settling time on the outcomes of the experiments was generally very limited. This is exemplified in figure\,\ref{fig:settling_time}\,({\it a,\,b\/}), presenting two representative trajectories, taken after waiting for 8 minutes and for more than 6 hours, respectively. Visually, there is no discernible difference between the two. For a more quantitative analysis, we present in figure\,\ref{fig:settling_time}\,({\it c,\,d,\,e\/}) results for the oscillation frequency $f$ and amplitude $\hat{a}$, as well as for the particle drag coefficient as a function of the waiting time. Both $f$ and $\hat{a}$ appear insensitive to the waiting time with differences within the range of the error bars, which primarily reflect the period-to-period variations that occur in these experiments. There is a slight initial decrease for $C_d$ with longer waiting times. However, around a waiting time of 10 minutes the curve levels off and $C_d$ stays constant for even longer waiting times. In summary, we therefore conclude that with waiting times $>8$ minutes our results presented in \S\,\ref{sec:Results} are not affected by residual disturbances in any significant manner.

An interesting observation in our experiments was that close to particle release, there was very little rotation of the particle and it appeared to rise more vertically compared to higher up in the measurement section of the water tank. The puzzling differences between the present results and those of \citet{Horowitz:2010} might therefore be related to a relatively long transient period, in which the spheres build up rotational kinetic energy. This is in line with a transient over 60--80$D$ in the present study, whereas the particles only travelled for $\approx 10D$ in \citet{Horowitz:2010} before entering the measurement section.

\section{Conclusion}\label{sec:Conclusion}
The systematic study performed here showed mixed results regarding the relevance of MOI variations for free rising spheres. On the one side, we identified a dependence of the drag coefficient on the MOI in the 3D chaotic regime. Namely, we found that $C_d$ increases with decreasing $I^*\Gamma$. This is accompanied by a trend of stronger rotations at lower $I^*\Gamma$ and therefore in line with findings of \citet{will2020rising}, who suggested a correlation between rotation rate and drag. Furthermore, our results indicate an increasing amplitude of horizontal path oscillations at lower $I^*\Gamma$, which confirms similar qualitative observations in \citet{Mathai2018}. Finally, at lower values of the MOI there is a stronger alignment between the rotation vector and the acceleration normal to the path. This indicates a more dominant rotational-translational coupling in this case and leads to fluttering-type trajectories.

On the other hand, the effect of MOI variations was limited to altering parameters within the 3D chaotic rise mode. It did not lead to regime changes in the overall behaviour, not even in case of turbulent flow, for which a transition from flutter to tumble has been reported previously \citep{Mathai2018}. The variation, e.g\ in $C_d$, induced by variations of $I^*\Gamma$ is hence limited and is not adequate to explain the scatter in the literature data. It appears that the rotational dynamics induced by centre of mass offsets, which can lead to drastic changes in the drag \citep{will2020rising}, play a more important role in this context.

Apart from the effect of the MOI, our results are also useful in revisiting the broader studied $\Gamma$-dependence, which together with $Re$ is the dominant parameter in governing the rise mode. Our analysis revealed the existence of a spiralling mode for $\Gamma \lessapprox 0.42$ for which the drag is significantly elevated ($C_d \approx 0.69$) similar to that found by \citet{Auguste2018} at lower $Ga$. Since neither the lateral amplitude nor the rotation rate excel similarly in this regime, it appears likely that the higher drag originates from a change in the wake structure.
3D chaotic motion was found for $\Gamma > 0.52$ with $C_d \approx 0.47$ (and the $I^*\Gamma$ dependence outlined above). This drag value is close to that of a fixed sphere at comparable $Re$ and matches also the results of \citet{preukschat1962} in the same range of density ratios ($\Gamma$) and the cases in \citet{Horowitz:2010} for $\Gamma \geq 0.61$. 
\citet{Horowitz:2010} also identified a density ratio triggered regime transition consistent with our observations, albeit with a different critical density ratio $\Gamma_{crit} \approx 0.61$ in the same $Ga$ regime. Other relevant differences between the present results and those of \citet{Horowitz:2010} relate to the existence of a rectilinear regime at larger $\Gamma$, which --- potentially due to a longer initial transient (see \S\,\ref{sec:waiting_time})--- is not observed in the present study. Based on varying the waiting time between experiments, we could rule out residual disturbances as a cause for this discrepancy.
In this regard our results appear consistent with \citet{preukschat1962}, who report a gradually decreasing but finite amplitude as $\Gamma \to 1$, in good agreement with the $I^*\Gamma$ dependence established here. Finally, even though the spiralling trajectories found at $\Gamma \lessapprox 0.42$ are in line with \citet{Karamanev:1992} and \citet{Auguste2018} at lower $Ga$, they are at odds with the 2D zigzag reported in \citet{Horowitz:2010} in this same parameter range.

\section*{Acknowledgments}
We thank Varghese Mathai, Detlef Lohse and Chao Sun fur helpful discussions. This work was supported by the Netherlands Organisation for Scientific Research (NWO) under VIDI Grant No. 13477. This project has received funding from the European Research Council (ERC) under the European Union’s Horizon 2020 research and innovation programme (grant agreement No. 950111  BU-PACT)
 
 \section*{Declaration of Interests}
The authors report no conflict of interest.

\renewcommand{\arraystretch}{1.1}

\appendix
\section{Tabulated experimental results}
\begin{longtable}{cccc|ccccccc}
$Ga$   & $\Gamma$ & $I^*$ & $D$(mm) & $\langle v_z \rangle_n$(m/s) & $\dfrac{\langle a_z \rangle_{rms}}{|1-\Gamma|g}$ & $Re$ & $C_d$ & Str & $\hat{a}/D$ & $\dfrac{\langle || \boldsymbol{\omega} || \rangle_n D}{V_b}$ \\
\hhline{====|=======}
\endhead
 \multicolumn{11}{c}{$0.37 < \Gamma < 0.42$ } \\ \hline \rowcolor{Grey}
5014 & 0.393    & 0.844 & 16.20    & 0.427                   & 0.104                                   & 6898 & 0.708 & 0.092 & 0.735 & 0.179                                                 \\ 
5056 & 0.377    & 0.819 & 16.15    & 0.434                   & 0.127                                   & 6983 & 0.705 & 0.092 & 0.730 & 0.179                                                 \\ \rowcolor{Grey}
4978 & 0.402    & 1.140 & 16.20    & 0.435                   & 0.095                                   & 7018 & 0.679 & 0.092 & 0.676 & 0.137                                                 \\ 
4960 & 0.406    & 1.147 & 16.20    & 0.449                   & 0.091                                   & 7245 & 0.633 & 0.088 & 0.639 & 0.158                                                 \\ \rowcolor{Grey}
4958 & 0.406    & 1.226 & 16.20    & 0.432                   & 0.113                                   & 6970 & 0.683 & 0.091 & 0.705 & 0.208                                                 \\ 
4956 & 0.412    & 1.460 & 16.25    & 0.419                   & 0.090                                   & 6786 & 0.717 & 0.095 & 0.670 & 0.155                                                 \\\hline
 \multicolumn{11}{c}{$0.52 < \Gamma < 0.57$ } \\ \hline \rowcolor{Grey}
4922 & 0.524    & 0.619 & 17.35    & 0.453                   & 0.062                                   & 7826 & 0.529 & 0.077 & 0.575 & 0.158                                                 \\ 
4890 & 0.530    & 0.620 & 17.35    & 0.454                   & 0.066                                   & 7842 & 0.521 & 0.061 & 0.752 & 0.240                                                 \\ \rowcolor{Grey}
4852 & 0.537    & 0.635 & 17.35    & 0.468                   & 0.041                                   & 8090 & 0.480 & 0.058 & 0.698 & 0.194                                                 \\ 
4884 & 0.531    & 0.794 & 17.35    & 0.439                   & 0.079                                   & 7595 & 0.555 & 0.064 & 0.771 & 0.309                                                 \\ \rowcolor{Grey}
4872 & 0.541    & 0.823 & 17.45    & 0.468                   & 0.043                                   & 8131 & 0.480 & 0.050 & 0.716 & 0.204                                                 \\
4851 & 0.541    & 0.946 & 17.40    & 0.451                   & 0.053                                   & 7814 & 0.517 & 0.060 & 0.658 & 0.243                                                 \\ \rowcolor{Grey}
4716 & 0.559    & 1.151 & 17.30    & 0.446                   & 0.053                                   & 7690 & 0.506 & 0.067 & 0.503 & 0.210                                                 \\ 
4788 & 0.553    & 1.149 & 17.40    & 0.453                   & 0.039                                   & 7860 & 0.497 & 0.064 & 0.516 & 0.194                                                 \\ \rowcolor{Grey}
4774 & 0.556    & 1.263 & 17.40    & 0.475                   & 0.026                                   & 8237 & 0.448 & 0.057 & 0.440 & 0.136                                                 \\ 
4699 & 0.562    & 1.259 & 17.30    & 0.469                   & 0.025                                   & 8088 & 0.451 & 0.063 & 0.380 & 0.137                                                 \\ \rowcolor{Grey}
4759 & 0.566    & 1.379 & 17.50    & 0.452                   & 0.037                                   & 7880 & 0.487 & 0.060 & 0.474 & 0.208                                                 \\ 
4794 & 0.556    & 1.379 & 17.45    & 0.465                   & 0.034                                   & 8081 & 0.470 & 0.045 & 0.659 & 0.198                                                 \\\hline
 \multicolumn{11}{c}{$0.66 < \Gamma < 0.70$ } \\ \hline \rowcolor{Grey}
4801 & 0.666    & 0.537 & 19.20    & 0.411                   & 0.061                                   & 7865 & 0.498 & 0.062 & 0.621 & 0.234                                                 \\
4814 & 0.669    & 0.698 & 19.30    & 0.415                   & 0.039                                   & 7987 & 0.487 & 0.049 & 0.626 & 0.193                                                 \\  \rowcolor{Grey}
4873 & 0.661    & 0.693 & 19.30    & 0.432                   & 0.015                                   & 8306 & 0.459 & 0.058 & 0.304 & 0.071                                                 \\ 
4834 & 0.666    & 0.827 & 19.30    & 0.426                   & 0.031                                   & 8191 & 0.465 & 0.045 & 0.587 & 0.156                                                 \\  \rowcolor{Grey}
4830 & 0.667    & 0.827 & 19.30    & 0.421                   & 0.030                                   & 8088 & 0.476 & 0.043 & 0.762 & 0.177                                                 \\ 
4739 & 0.682    & 1.027 & 19.35    & 0.414                   & 0.030                                   & 7991 & 0.471 & 0.045 & 0.542 & 0.154                                                 \\  \rowcolor{Grey}
4713 & 0.685    & 1.191 & 19.35    & 0.416                   & 0.030                                   & 8017 & 0.462 & 0.047 & 0.439 & 0.145                                                 \\
4763 & 0.681    & 1.197 & 19.40    & 0.416                   & 0.021                                   & 8048 & 0.468 & 0.043 & 0.502 & 0.149                                                 \\  \rowcolor{Grey}
4635 & 0.696    & 1.246 & 19.35    & 0.416                   & 0.024                                   & 8030 & 0.445 & 0.046 & 0.439 & 0.116                                                 \\ 
4704 & 0.686    & 1.247 & 19.35    & 0.412                   & 0.028                                   & 7941 & 0.468 & 0.039 & 0.611 & 0.179                                                 \\\hline
 \multicolumn{11}{c}{$0.79 < \Gamma < 0.85$ } \\ \hline \rowcolor{Grey}
4720 & 0.794    & 0.607 & 22.30    & 0.362                   & 0.034                                   & 8043 & 0.460 & 0.047 & 0.810 & 0.172                                                 \\ 
4735 & 0.792    & 0.605 & 22.30    & 0.361                   & 0.025                                   & 8013 & 0.466 & 0.042 & 0.610 & 0.156                                                 \\ \rowcolor{Grey}
4556 & 0.808    & 0.733 & 22.30    & 0.356                   & 0.022                                   & 7904 & 0.443 & 0.037 & 0.649 & 0.159                                                 \\
4609 & 0.806    & 0.738 & 22.40    & 0.353                   & 0.022                                   & 7889 & 0.456 & 0.042 & 0.554 & 0.109                                                 \\ \rowcolor{Grey}
4572 & 0.808    & 0.830 & 22.35    & 0.348                   & 0.021                                   & 7755 & 0.465 & 0.047 & 0.446 & 0.125                                                 \\ 
4584 & 0.808    & 0.911 & 22.40    & 0.352                   & 0.020                                   & 7846 & 0.456 & 0.043 & 0.452 & 0.121                                                 \\ \rowcolor{Grey}
4558 & 0.810    & 0.913 & 22.40    & 0.347                   & 0.025                                   & 7745 & 0.462 & 0.042 & 0.520 & 0.177                                                 \\ 
4380 & 0.825    & 1.035 & 22.40    & 0.343                   & 0.017                                   & 7655 & 0.437 & 0.042 & 0.431 & 0.118                                                 \\ \rowcolor{Grey}
4440 & 0.820    & 1.034 & 22.40    & 0.337                   & 0.020                                   & 7520 & 0.465 & 0.045 & 0.467 & 0.134                                                 \\
4314 & 0.830    & 1.141 & 22.40    & 0.333                   & 0.020                                   & 7433 & 0.450 & 0.041 & 0.515 & 0.136                                                 \\ \rowcolor{Grey}
4395 & 0.825    & 1.142 & 22.45    & 0.329                   & 0.022                                   & 7370 & 0.475 & 0.037 & 0.709 & 0.187                                                 \\ 
4300 & 0.832    & 1.205 & 22.45    & 0.325                   & 0.023                                   & 7280 & 0.466 & 0.042 & 0.540 & 0.149                                                 \\ \rowcolor{Grey}
4308 & 0.832    & 1.205 & 22.45    & 0.329                   & 0.020                                   & 7359 & 0.458 & 0.035 & 0.391 & 0.126                                                 \\ 
4248 & 0.835    & 1.220 & 22.40    & 0.325                   & 0.022                                   & 7253 & 0.458 & 0.044 & 0.466 & 0.128                                                 \\ \rowcolor{Grey}
4138 & 0.843    & 1.220 & 22.35    & 0.332                   & 0.025                                   & 7391 & 0.418 & 0.037 & 0.667 & 0.157                                                 \\\hline
 \multicolumn{11}{c}{$0.90 < \Gamma < 0.97$ } \\ \hline \rowcolor{Grey}
4474 & 0.918    & 0.519 & 29.25    & 0.239                   & 0.056                                   & 6953 & 0.558 & 0.059 & 0.915 & 0.287                                                 \\
4777 & 0.906    & 0.680 & 29.25    & 0.274                   & 0.024                                   & 7976 & 0.479 & 0.047 & 0.583 & 0.180                                                 \\ \rowcolor{Grey}
4705 & 0.909    & 0.682 & 29.25    & 0.276                   & 0.020                                   & 8038 & 0.457 & 0.043 & 0.533 & 0.116                                                 \\
4062 & 0.933    & 0.831 & 29.35    & 0.217                   & 0.021                                   & 6340 & 0.549 & 0.046 & 0.550 & 0.161                                                 \\ \rowcolor{Grey}
4269 & 0.926    & 0.829 & 29.40    & 0.227                   & 0.019                                   & 6636 & 0.553 & 0.044 & 0.504 & 0.127                                                 \\
4799 & 0.907    & 0.822 & 29.40    & 0.263                   & 0.027                                   & 7707 & 0.518 & 0.042 & 0.770 & 0.155                                                 \\ \rowcolor{Grey}
4688 & 0.911    & 0.822 & 29.35    & 0.266                   & 0.018                                   & 7784 & 0.484 & 0.051 & 0.340 & 0.068                                                 \\
4559 & 0.915    & 0.923 & 29.30    & 0.266                   & 0.017                                   & 7780 & 0.458 & 0.055 & 0.317 & 0.094                                                 \\ \rowcolor{Grey}
4701 & 0.910    & 0.923 & 29.35    & 0.265                   & 0.015                                   & 7765 & 0.489 & 0.043 & 0.464 & 0.118                                                 \\
4678 & 0.911    & 0.923 & 29.35    & 0.267                   & 0.014                                   & 7806 & 0.480 & 0.042 & 0.629 & 0.119                                                 \\ \rowcolor{Grey}
4493 & 0.918    & 1.000 & 29.30    & 0.266                   & 0.016                                   & 7775 & 0.445 & 0.049 & 0.334 & 0.098                                                 \\
4608 & 0.913    & 0.999 & 29.30    & 0.276                   & 0.019                                   & 8059 & 0.436 & 0.046 & 0.453 & 0.117                                                 \\ \rowcolor{Grey}
4474 & 0.918    & 1.000 & 29.30    & 0.269                   & 0.017                                   & 7852 & 0.433 & 0.047 & 0.403 & 0.081                                                 \\
3286 & 0.956    & 1.002 & 29.30    & 0.202                   & 0.013                                   & 5902 & 0.413 & 0.047 & 0.279 & 0.101                                                 \\ \rowcolor{Grey}
3453 & 0.952    & 1.001 & 29.35    & 0.200                   & 0.010                                   & 5849 & 0.465 & 0.042 & 0.266 & 0.095                                                 \\
4361 & 0.923    & 1.058 & 29.35    & 0.253                   & 0.017                                   & 7394 & 0.464 & 0.042 & 0.622 & 0.141                                                 \\ \rowcolor{Grey}
4528 & 0.917    & 1.058 & 29.40    & 0.258                   & 0.015                                   & 7553 & 0.479 & 0.045 & 0.410 & 0.117                                                 \\
4540 & 0.917    & 1.058 & 29.40    & 0.255                   & 0.013                                   & 7478 & 0.492 & 0.045 & 0.406 & 0.093                                                 \\ \rowcolor{Grey}
3055 & 0.962    & 1.050 & 29.35    & 0.175                   & 0.011                                   & 5108 & 0.478 & 0.042 & 0.408 & 0.108                                                 \\
3320 & 0.955    & 1.051 & 29.40    & 0.180                   & 0.015                                   & 5260 & 0.532 & 0.040 & 0.521 & 0.138                                                 \\ \rowcolor{Grey}
4298 & 0.925    & 1.113 & 29.35    & 0.255                   & 0.013                                   & 7469 & 0.442 & 0.054 & 0.251 & 0.117                                                 \\
4504 & 0.918    & 1.113 & 29.40    & 0.249                   & 0.017                                   & 7282 & 0.510 & 0.046 & 0.458 & 0.155                                                 \\ \rowcolor{Grey}
4416 & 0.922    & 1.115 & 29.45    & 0.252                   & 0.019                                   & 7381 & 0.477 & 0.042 & 0.577 & 0.136                                                 \\
4114 & 0.932    & 1.178 & 29.40    & 0.250                   & 0.016                                   & 7319 & 0.422 & 0.051 & 0.311 & 0.114                                                 \\ \rowcolor{Grey}
4290 & 0.926    & 1.178 & 29.40    & 0.245                   & 0.018                                   & 7183 & 0.476 & 0.053 & 0.290 & 0.113                                                 \\
4090 & 0.932    & 1.177 & 29.35    & 0.250                   & 0.014                                   & 7301 & 0.419 & 0.051 & 0.256 & 0.094                                                 \\\hline
\caption{Tabulated particle parameters and results.}
	\label{tab:prop}
\end{longtable}

\bibliographystyle{jfm}
\bibliography{Citations_MoI}

\end{document}